\documentclass[twocolumn, switch]{article} % Method A for two-column formatting

\usepackage{preprint}

\usepackage{amsmath, amsthm, amssymb, amsfonts}

\usepackage[numbers,square]{natbib}
\usepackage[utf8]{inputenc} % allow utf-8 input
\usepackage[T1]{fontenc}    % use 8-bit T1 fonts
\usepackage{xcolor}         % colors for hyperlinks
\usepackage[colorlinks = true,
            linkcolor = purple,
            urlcolor  = blue,
            citecolor = cyan,
            anchorcolor = black]{hyperref}  % Color links to references, figures, etc.
\usepackage{nicefrac}       % compact symbols for 1/2, etc.
\usepackage{microtype}      % microtypography
\usepackage{lineno}         % Line numbers
\usepackage{float}          % Allows for figures within multicol

\usepackage{lipsum}         % Filler text
\usepackage{tabularx}
\usepackage{graphicx}
\usepackage[export]{adjustbox}

\usepackage{newfloat}
\DeclareFloatingEnvironment[name={Supplementary Figure}]{suppfigure}
\usepackage{sidecap}
\sidecaptionvpos{figure}{c}

\usepackage{titlesec}
\titlespacing\section{0pt}{12pt plus 3pt minus 3pt}{1pt plus 1pt minus 1pt}
\titlespacing\subsection{0pt}{10pt plus 3pt minus 3pt}{1pt plus 1pt minus 1pt}
\titlespacing\subsubsection{0pt}{8pt plus 3pt minus 3pt}{1pt plus 1pt minus 1pt}

\title{Automatic normal orientation in point clouds of building interiors}

\usepackage{authblk}

\author[1]{Sebastian Ochmann}
\author[1]{Reinhard Klein}
\affil[1]{Institute of Computer Science II, University of Bonn, Germany}

\begin{document}

\twocolumn[ % Method A for two-column formatting
    \begin{@twocolumnfalse} % Method A for two-column formatting
  
\maketitle

\begin{abstract}
Orienting surface normals correctly and consistently is a fundamental problem in geometry processing.
Applications such as visualization, feature detection, and geometry reconstruction often rely on the availability of correctly oriented normals.
Many existing approaches for automatic orientation of normals on meshes or point clouds make severe assumptions on the input data or the topology of the underlying object which are not applicable to real-world measurements of urban scenes.
In contrast, our approach is specifically tailored to the challenging case of unstructured indoor point cloud scans of multi-story, multi-room buildings.
We evaluate the correctness and speed of our approach on multiple real-world point cloud datasets.

\end{abstract}
%\keywords{Point Clouds \and Normal Orientation} % (optional)
\vspace{0.35cm}
\end{@twocolumnfalse} % Method A for two-column formatting
] % Method A for two-column formatting

%\begin{multicols}{2} % Method B for two-column formatting (doesn't play well with line numbers), comment out if using method A

%%%%%%%%%%%%%%%  Main text   %%%%%%%%%%%%%%%
% \linenumbers

\newcommand{\figBounces}{
\begin{figure*}
    \includegraphics[width=\textwidth]{{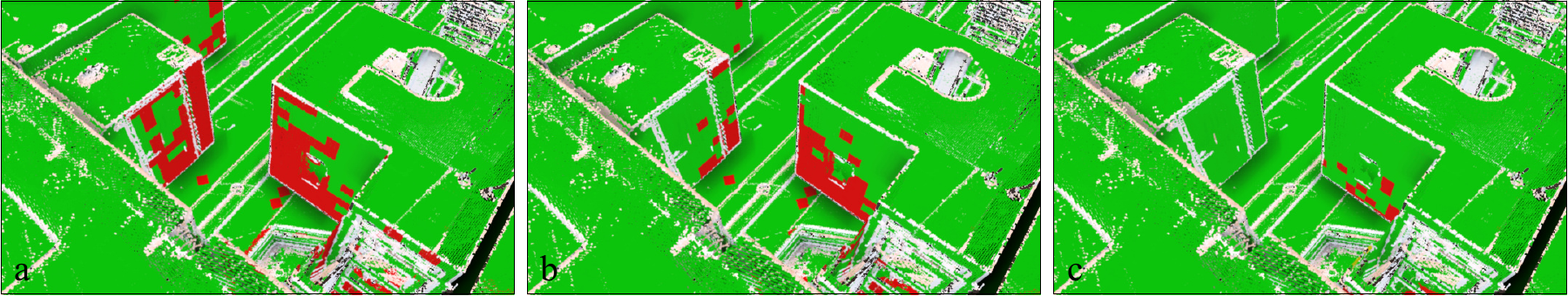}}
    \caption{The effect of using path tracing with varying number of bounces. The images show the orientation correctness (green = correct, red = incorrect) after the path tracing phase, but before applying surface consistency. Note how the region in the center of the images is occluded by the surrounding rooms, and also by the story below. (a) Using only a single bounce leads to uncertainty in occluded regions. (b) This effect is slightly improved when increasing the number of bounces to two. (c) Eight bounces as used in our approach leads to results which are easily corrected by the subsequent surface consistency voting.}
    \label{fig:bounces}
\end{figure*}
}

\newcommand{\figClassification}{
\begin{figure}
    \includegraphics[width=\columnwidth]{{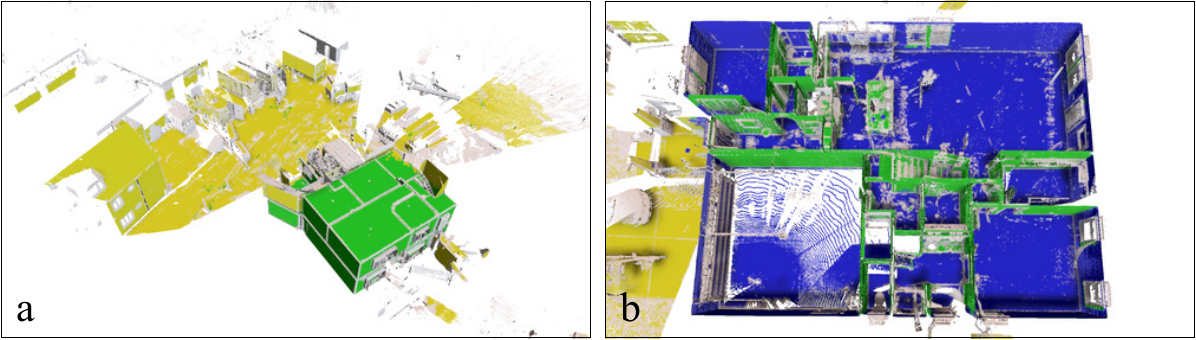}}
    \caption{Overview of the classification of the point cloud done as part of our orientation algorithm. (a) Classification into outside (yellow) and non-outside (green) parts. Points not on patches are colored gray. (b) Points are furthermore classified as belonging to surfaces between room interior and outside (blue) and between neighboring rooms (green).}
    \label{fig:classification}
\end{figure}
}

\newcommand{\figFacade}{
\begin{figure}
    \includegraphics[width=\columnwidth]{{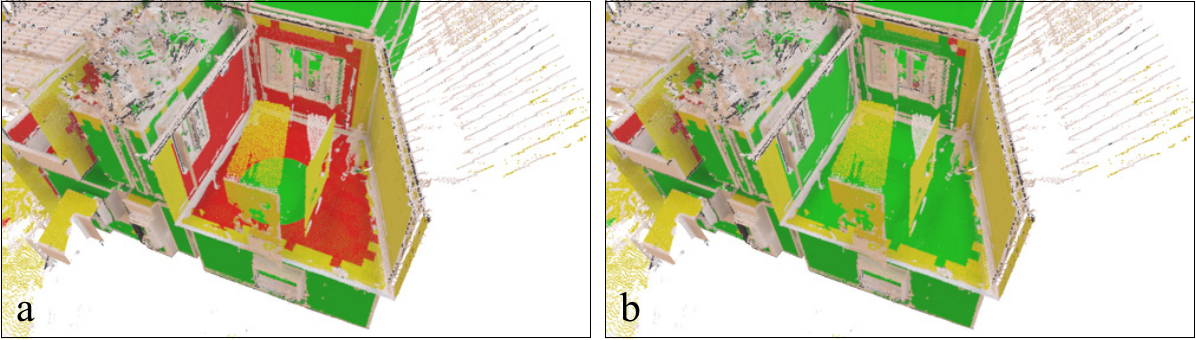}}
    \caption{(a) The path tracing phase (Section \ref{subsec:pathtracing}) may orient points on fa\c{c}ade parts incorrectly. (b) The second phase (Section \ref{subsec:facade}) attempts to correct these cases by flipping the respective normals.}
    \label{fig:facade}
\end{figure}
}

\newcommand{\figFacadeFail}{
\begin{figure}
    \includegraphics[width=\columnwidth]{{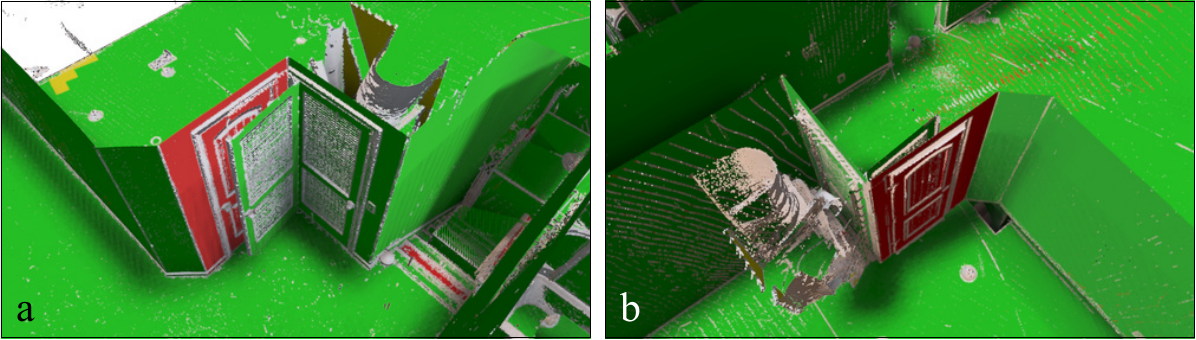}}
    \caption{A failure case of the fa\c{c}ade correction phase (Section \ref{subsec:facade}. (a) and (b) are different views of the same scene. The surface highlighted red has incorrect orientation since the opened, almost parallel door was interpreted as a wall surface.}
    \label{fig:facade_fail}
\end{figure}
}

\newcommand{\evalimg}[2]{
    \includegraphics[width=4.1cm]{images/#1/screenshot_#2.jpg}
}

\newcommand{\evalimgline}[1]{
    \evalimg{Gormsgade1}   {#1} &
    \evalimg{Gormsgade3}   {#1} &
    \evalimg{LTU}          {#1} &
    \evalimg{SmallBuilding}{#1} \\
}

\newcommand{\tblEvaluation}{

\begin{table*}

\small

\begin{tabularx}{\textwidth}{|X|X|X|X|}
    \hline
%   Gormsgade1                  & Gormsgade3                    & LTU                           & SmallBuilding
    Dataset 1                   & Dataset 2                     & Dataset 3                     & Dataset 4                     \\
    \hline
    \textbf{Statistics}         & \textbf{Statistics}           & \textbf{Statistics}           & \textbf{Statistics}           \\
    \# points / scans: 5,151,388 / 21        & \# points / scans: 7,688,111 / 29      & \# points / scans: 12,409,443 / 13     & \# points / scans: 34,964,707 / 39    \\
    points on patches: 73\%     & points on patches: 71\%       & points on patches: 73\%       & points on patches: 83\%       \\
    points non-outside: 68\%    & points non-outside: 59\%      & points non-outside: 67\%      & points non-outside: 80\%      \\
    \hline
    \textbf{Plane detection}    & \textbf{Plane detection}      & \textbf{Plane detection}      & \textbf{Plane detection}      \\
    \# planes: 228              & \# planes: 320                & \# planes: 322                & \# planes: 556                \\
    Runtime: 29,320ms           & Runtime: 94,468ms             & Runtime: 114,929ms            & Runtime: 228,903ms            \\
    \hline
    \textbf{Correctness}        & \textbf{Correctness}          & \textbf{Correctness}          & \textbf{Correctness}          \\
    Phase 1A: 97.94\%           & Phase 1A: 96.14\%             & Phase 1A: 98.08\%             & Phase 1A: 98.64\%             \\
    Phase 1B: 98.75\%           & Phase 1B: 97.11\%             & Phase 1B: 98.05\%             & Phase 1B: 98.98\%             \\
    Phase 2A: 98.09\%           & Phase 2A: 98.20\%             & Phase 2A: 97.53\%             & Phase 2A: 98.45\%             \\
    Phase 2B: \textbf{98.82\%}  & Phase 2B: \textbf{98.75\%}    & Phase 2B: \textbf{99.42\%}    & Phase 2B: \textbf{98.68\%}    \\
    \hline
    \textbf{Runtime}            & \textbf{Runtime}              & \textbf{Runtime}              & \textbf{Runtime}              \\
    Phase 1A: 1885ms            & Phase 1A: 3751ms              & Phase 1A: 3963ms              & Phase 1A: 4142ms              \\
    Phase 1B: 186ms             & Phase 1B: 780ms               & Phase 1B: 582ms               & Phase 1B: 1462ms              \\
    Phase 2A: 102ms             & Phase 2A: 212ms               & Phase 2A: 163ms               & Phase 2A: 232ms               \\
    Phase 2B: 186ms             & Phase 2B: 756ms               & Phase 2B: 581ms               & Phase 2B: 1461ms              \\
    \hline
    \evalimgline{6}
    \evalimgline{7}
    \evalimgline{11}
    \hline
    \evalimgline{12}
    \evalimgline{13}
    \evalimgline{17}
    \hline
\end{tabularx}

\caption{Evaluation results on real-world datasets. ``Statistics'': Number of points and scans of the input data, the ratio of points on patches, and on patches that are not discarded as outside. ``Plane detection'': The number of detected planes and the runtime of the shape detection algorithm. ``Correctness'': Ratio of correctly oriented points to the total number of points on non-outside patches. The percentage after phase 2B is the final result. ``Runtime'': The computation time for the individual steps of our approach. The images show cross sections of the upper and lower story of each dataset. For each story, the unlabeled point cloud is shown, the classification into interior/exterior/outside surfaces (green/blue/yellow), and whether the orientation is correct/incorrect (green/red).}
\label{tbl:evaluation}

\end{table*}

}

\section{Introduction}
\label{sec:introduction}

For many applications in computer graphics and related domains, surface normals are an important property of 3D point cloud or mesh data.
While normal \emph{directions} can usually be estimated sufficiently well by analyzing local surface properties using e.g.\ principal component analysis (PCA) in case of point clouds, automatically determining the correct normal \emph{orientation}, i.e.\ the sign of the normal vectors, generally is a much harder problem.
In particular for mesh data, there exists a wide variety of approaches based on principles such as voting, visibility, propagation, and optimization.
Since point clouds are increasingly used as a means for representing various kinds of objects and scenes in fields like architecture, design, archaeology, and cultural heritage, methods working directly on point clouds have also received attention.
A particularly important and discriminating aspect of any method is the severity of the assumptions made on the input and the particular geometry represented by the data.
Regarding the input data, the assumptions range from very restrictive such as watertight, connected meshes, to unconnected polygon soups, possibly with missing parts.
Point clouds pose additional challenges for certain kinds of methods based on connectivity for propagation, or surfaces for performing ray casting against, since such information is not directly available from the data.
With respect to the class of the underlying object or scene, some methods make the assumption that the object itself is a closed 2-manifold.
While this assumption simplifies the task of distinguishing between inside and outside space, many kinds of larger-scale datasets such as 3D urban environments do not fulfill this requirement.

Our work targets the challenging task of automatically determining normal orientations in completely unstructured 3D point cloud datasets of building interiors with multiple stories and rooms.
We are specifically interested in the main structure of the building consisting of floor, ceiling, and wall surfaces.
This information is an important prerequisite for e.g.\ reconstruction tasks aiming at automatic generation of higher-level 3D models from point cloud data.
Clearly, knowledge about correct surface orientation in previously unstructured data greatly helps to distinguish between room interior, interior of wall volumes, and outside area.
Point cloud scans of building interiors, possibly including parts of exterior fa\c{c}ade and parts of the outside area scanned through windows, pose two particular challenges.
First, such datasets usually consist of millions of points and cover a relatively large area which requires efficient means of processing them.
Second, the constellation of rooms within a building can be quite complex, yielding a much more intricate surface topology than 2-manifolds.

The proposed method for automatically orienting normals in point clouds of building interiors combines different ideas to provide efficient processing of real-world scans.
We first simplify the scene by detecting planes in the point cloud and subsequently working on surface patches instead of individual points.
One advantage of working on patches instead of individual points is the drastically reduced computational complexity.
In addition, the surface representation enables us to employ a specifically tailored path tracing approach to estimate which side of each patch is probably room interior, wall, or outside area.
While visibility information is exploited by several algorithms, our method not only takes direct visibility into account but also higher-order visibility through multiple ray bounces.
Using this initial, per-patch estimation, we then vote for a global orientation for each surface to increase the robustness of the estimation.
Finally, the determined surface orientations are used to flip the normal orientations of the points belonging to the respective surface.
Our approach is evaluated on multiple real-world datasets for which ground truth normal orientations for comparison are available by means of known scanner positions.

% Mention confidence estimation?
% Comparison against off-the-shelf-methods?

In summary, the contribution of our approach is a fast and fully automatic normal orientation estimation for the challenging scenario of indoor building scans without strong assumptions on the input data.
The results of our method can greatly facilitate tasks such as reconstruction of 3D models from point clouds which require knowledge about the orientation of surfaces of the main building structure such as floors, ceilings, and walls.

\section{Related Work}
\label{sec:relatedwork}

A classical propagation-based approach for orienting normals of point sets is described by Hoppe et al.\ \cite{Hoppe_1992_Reconstruction}. It derives a consistent orientation of tangent planes for data points by means of solving an optimization problem on the Riemannian graph of the points with edge weights proportional to the normal direction deviation between neighboring points. The method is only applicable for densely sampled, closed surfaces and may fail at sharp creases.
König et al.\ \cite{Koenig_2009_Consistent} base their method on the method by Hoppe et al.\ \cite{Hoppe_1992_Reconstruction} but propose a new unreliability cost for traversing the Riemannian graph based on Hermite curves.
One recent point cloud based approach by Schertler et al.\ \cite{Schertler_2017_Globally} generalizes propagation as a graph-based energy minimization problem. To this end, the graph-based idea by Hoppe et al.\ \cite{Hoppe_1992_Reconstruction} is reformulated to a maximum-likelihood problem on a Markov random field. They also propose to use the streaming approach by Pajarola \cite{Pajarola_2005_Stream} to perform out-of-core processing.

The volumetric approach to solid inside/outside classification of polygonal data by Murali et al.\ \cite{Murali_1997_Solid} is based on a partitioning of space into polyhedral cells on which a consistent classification is derived by optimization.
Xie et al.\ \cite{Xie_2004_Noisy} segment an input point cloud into so-called mono-oriented regions through an active contour method. Subsequently, a consistent inside/outside partitioning is achieved by means of a voting algorithm.
The approach by Mello et al.\ \cite{Mello_2003_InOut} constructs an adaptively subdivided tetrahedral decomposition from an input point cloud for which a consistent labeling as inside/outside over all tetrahedra is determined by means of a simulated annealing approach.
Alliez et al.\ \cite{Alliez_2007_Voronoi} present a variational framework for combined normal direction and orientation estimation as part of their surface reconstruction approach. They first compute a tensor field using a Voronoi diagram of the input point set and derive a best-fitting isosurface by solving a generalized eigenvalue problem.
Another variational approach which finds normal directions and orientations simultaneously is presented by Wang et al.\ \cite{Wang_2012_Variational}.
Liu et al.\ \cite{Liu_2010_Orienting} transfer an input point cloud to a coarse triangulated mesh in order to determine normal orientations on this mesh representing the underlying topology. This information is subsequently used to orient normals on the original point set.

An approach which employs stochastic ray voting is presented by Mullen et al.\ \cite{Mullen_2010_Signing}. An unsigned distance function is first estimated on a 3D Delaunay triangulation of an input point set. Initial estimates for the sign of the distance function are obtained by means of ray shooting and testing for intersections with an $\varepsilon$-band of the unsigned function which is then smoothed and propagated.
Borodin et al.\ \cite{Borodin_2004_Consistent} combine a proximity- and visibility based approach to orient polygons in meshes. A connectivity graph between patches of the model is constructed in which each patch has two visibility coefficients which encode how much of the two sides of the patch is visible from outside. To achieve this, one of the proposed methods is a ray casting approach similar to ours. However, the assumption is that most of the object's surface is visible from outside the model.
Takayama et al.\ \cite{Takayama_2014_Raycasting} also employ a ray casting based approach to orient facets in polygon meshes. They cast rays in both directions of facets to determine where outside space is located. For inner facets, they attempt to determine which side of the facet has more free space than the other which is similar to our idea for inner walls. Since this method may fail in cavities, they propose an alternative method based on intersection parity which is prone to modeling errors.
In contrast, we employ path tracing with multiple bounces to deal with cavities in the scene.
One method that implicitly considers ray paths to propagate inside/outside classifications in triangular meshes is presented by Zhou et al.\ \cite{Zhou_2008_Visibility}. Based on point samples on triangles, a weighted visibility graph is constructed between the points whose nodes are classified as inside or outside using graph cut.

\section{Method}
\label{sec:method}

The input of our approach is a set of points in $\mathbb{R}^3$ mainly representing the interior of a building, possibly with some parts of exterior fa\c{c}ade and parts of outside area.
If (unoriented) normals are not given in the data, they are estimated by means of local principal component analysis (PCA) for each point.

\subsection{Plane detection and patch generation}

We first detect planes in the point cloud data to obtain a simplified and more structured representation of the scene.
Detection of primitive shapes in point clouds is a well studied problem and any reasonable method can be applied.
We use the CGAL implementation \cite{CGAL-2018-Shapes} of the random sample consensus (RANSAC) method by Schnabel et al.\ \cite{Schnabel-2007-Primitives} for its efficiency and quality of the resulting shapes.
The rationale behind using planes is that the main structure of buildings can usually be represented well in a piecewise planar manner.
Note that other shapes such as spheres or cylinders are also supported by the detection algorithm and could in principle also be used for representing e.g.\ columns or curved walls.

For each of the detected planes, a relatively coarse 2D \emph{occupancy bitmap}, i.e.\ a uniform grid on the surface on which each cell or pixel may have the value $0$ or $1$, represents the support of the plane by the points constituting the plane. A pixel of the bitmap has the value $1$ if and only if at least one point is located within the pixel.
All pixels with value $1$ yield the set $P$ of patches which will be used in the following steps.
Each patch $p \in P$ originates from an original surface (i.e.\ plane) $s_p$, has a center position $c_p \in \mathbb{R}^3$ and an initial normal $\tilde{n}_p$ with arbitrary but fixed orientation.

\subsection{Orientation by path tracing}
\label{subsec:pathtracing}

Our first goal is to estimate an initial normal orientation for each individual patch.
Specifically, given a patch $p \in P$, there are two possible orientations for its normal, $\tilde{n}_p$ and $-\tilde{n}_p$.
For most points in the datasets we consider, we wish to select the one orientation which points towards the interior of a room.
Conversely, the normal should point away from outside area (in case the patch is part of a surface separating room interior and outside area), and away from the interior of wall, floor or ceiling structures (in case the surface separates neighboring rooms).

This classification task is formulated as a voting scheme based on path tracing.
Intuitively, for each patch, we trace a number of random paths into both hemispheres for the two possible orientations.
We then use the number of ray bounces as well as the path lengths to analyze two aspects.
First, we classify whether patches belong to interior or exterior walls, or are located completely outside of the building.
Second, we use this classification as well as the path lengths to flip the normal of each patch to the more likely correct orientation.
Finally, the reoriented patch normals vote for a normal orientation of whole surfaces.

\newcommand{\Lpos}{L_{p, \tilde{n}_p}}
\newcommand{\Lneg}{L_{p, -\tilde{n}_p}}
\newcommand{\Bpos}{B_{p, \tilde{n}_p}}
\newcommand{\Bneg}{B_{p, -\tilde{n}_p}}

We now formalize the approach.
All ray intersections are tested against the set of patches $P$.
Let us first consider a patch $p$ with center $c_p \in \mathbb{R}^3$ and one specific orientation $\tilde{n}_p$.
We cast $k$ rays $r_i$, $i \in \{1, \dots, k\}$, each with origin $c_p$ and a direction randomly sampled within a $120^\circ$ cone directed towards $\tilde{n}_p$.
In our experiments, $k = 50$.
If a ray $r_i$ with direction $d_{r_i}$ intersects with a patch $p'$, the ray is reflected into a sampled direction within a cone oriented towards the hemisphere of the incoming ray. The direction $n_H$ of the hemisphere is computed as
\[
n_H =
\begin{cases}
 \tilde{n}_{p'}, & \text{if } \langle \tilde{n}_{p'}, d_{r_i} \rangle < 0, \\
-\tilde{n}_{p'}, & \text{otherwise.}
\end{cases}
\]
Note that the normal $\tilde{n}_{p'}$ of the intersected patch $p'$ used for this computation is arbitrary but fixed.
In particular, the path tracing is invariant under the initial orientation of the patches.
We allow up to $b = 8$ ray bounces for each initial ray $r_i$.
If a ray does not hit any patch, the respective path is terminated at that point.
The result are $k$ ray paths, each with up to $b$ bounces for the considered patch $p$ and orientation $\tilde{n}_p$.

Let $l^{i,j}_{p, \tilde{n}_p}$ be the length of the $j$th segment along the $i$th path traced for patch $p$ and orientation $\tilde{n}_p$ (note that segments after termination of a ray are considered to have zero length).
We define the accumulated length $\Lpos$ as
\[
\Lpos = \sum_{i=1}^k \sum_{j=1}^b \log(1 + l^{i,j}_{p, \tilde{n}_p}).
\]
The rationale for taking the logarithm is to decrease the influence of particularly long segments while still distinguishing between short and medium-length segments.
Furthermore, let $b_i$ be the number of bounces of the $i$th path. We consider the average number of ray bounces $\Bpos$ over all $k$ paths
\[
\Bpos = \frac{1}{k} \sum_{i=1}^k b_i.
\]

Note that we analogously have $\Lneg$ and $\Bneg$ for the opposite direction.
We now define a classification function $C(p) : P \to \{in, ex, out\}$ of patch $p$ into interior, exterior, or outside as
\[
C(p) =
\begin{cases}
out & \text{if } (\Bpos < \tau) \text{ and } (\Bneg < \tau) \\
ex  & \text{if } (\Bpos < \tau) \text{ xor } (\Bneg < \tau) \\
in  & \text{otherwise,}
\end{cases}
\]
where $\tau$ is a threshold which was empirically chosen as $4$ in our experiments.

A patch $p$ with $C(p) = out$ is considered to be clutter outside of the building and subsequently ignored.
A patch with $C(p) = ex$ is considered to be part of a surface separating room interior from outside area. Its corrected normal orientation $\hat{n}_p$ is set to point away from the outside area, i.e.
\[
\hat{n}_p =
\begin{cases}
 \tilde{n}_p & \text{if } \Bneg < \tau \\
-\tilde{n}_p & \text{if } \Bpos < \tau.
\end{cases}
\]
A patch with $C(p) = in$ is considered to be part of a surface between neighboring rooms.
In this case, we assume that the orientation with the longer total path length points towards the room interior and we thus set the corrected normal orientation to
\[
\hat{n}_p =
\begin{cases}
 \tilde{n}_p & \text{if } L_{p, \tilde{n}_p} > L_{p, -\tilde{n}_p}, \\
-\tilde{n}_p & \text{otherwise.}
\end{cases}
\]

The orientation estimation up to this point was performed separately for each patch.
Assuming that all points of each of the originally detected planes share a common normal orientation, we can easily vote for an orientation using all patches belonging to a common plane.
Let $s$ be one of the detected planes with arbitrarily oriented normal $\tilde{n}_s$ and let $P_s = \{p \ \vert\ s_p = s\}$ be the set of patches originating from surface $s$.
For voting, we determine the value
\[
\theta_s = \sum_{p \in P_s} \text{sgn}\left( \langle \hat{n}_{p}, \tilde{n}_s \rangle \right),
\]
where $\text{sgn}(\cdot)$ is the standard signum function, and determine the corrected surface normal $n_s$ as
\[
n_s =
\begin{cases}
 \tilde{n}_s, & \text{if } \theta_s > 0, \\
-\tilde{n}_s, & \text{otherwise.}
\end{cases}
\]
Then all patch normals are flipped to point in the same direction as $n_s$.
For simplicity, we will still call this corrected normal $\hat{n}_p$ in the following.

\subsection{Correction for fa\c{c}ade parts}
\label{subsec:facade}

Surfaces belonging to exterior fa\c{c}ade are sometimes encountered in interior scans due to scanning through windows.
For such patches, the above estimation may erroneously prefer the direction pointing away from the outside area since ray paths towards the outside area are terminated quickly while rays towards the exterior wall of the building generate longer paths.
An example for such an erroneous estimation is shown in Figure \ref{fig:facade}.
To correct the orientation in these cases, we perform a second, simpler ray casting pass as follows.
For each patch $p$ with center $c_p$ and orientation $\hat{n}_p$ as estimated above, we cast $k$ rays $r_i$, $i \in \{1, \dots, k\}$, originating at $c_p$ with directions $d_i$ sampled in a cone oriented towards $\hat{n}_p$ without allowing ray bounces.
Let $p'_i$ be the patch which is hit by ray $r_i$.
We then consider the value of
\[
\phi_p = \sum_{i=1}^k \text{sgn}\left( \langle \hat{n}_{p'_i}, d_i \rangle \right).
\]
If $\phi_p > 0$, the estimated orientation $\hat{n}_p$ is probably incorrect since it points towards the back side of a surface of a room interior.
We thus define the corrected oriented normal $\overline{n}_p$ as
\[
\overline{n}_p =
\begin{cases}
 \hat{n}_p & \text{if } \phi_p < 0, \\
-\hat{n}_p & \text{otherwise.}
\end{cases}
\]

The patch normals are then again used to vote for a common normal orientation within each surface in the same way as described at the end of Section \ref{subsec:pathtracing}.

As a final step, the oriented normals of the patches are used to orient the normals of the original points of the point cloud which lie within the respective patch.

\section{Evaluation}
\label{sec:evaluation}

\tblEvaluation

\figBounces

\figClassification

\figFacade

\figFacadeFail

To test the correctness and runtime of our approach, we applied it to multiple real-world datasets with ground truth normal orientations.
Specifically, the datasets consist of multiple, registered scans with known scanner positions for each scan.
This allows us to flip normals towards the respective scanner positions to obtain the correct orientations.
In order to test our approach, we ignore the known orientation and scanner positions, and then compare our estimated orientations with the ground truth.
We also measure the runtime of the main processing steps.
Table \ref{tbl:evaluation} summarizes the results of our experiments which are further discussed below.

\subsection{Input data, planes, and patches}

The first part of Table \ref{tbl:evaluation} shows general statistics about the datasets such as number of points and scans.
Note that information about individual scans and scanner positions is only used for generating ground truth normal orientations.
It also lists the percentage of the total points which are part of detected planes (and thus belong to patches), and the percentage of the total points which are on patches that are \emph{not} classified as outside area (i.e.\ $C(p) \neq out$).
Note that it is exactly this set of non-outside points for which our algorithm estimates normal orientations, and that the correctness is measured with respect to this set of points.
The number of detected planes and the runtime of plane detection using the RANSAC implementation of the CGAL library \cite{CGAL-2018-Shapes} is also listed in the table.

\subsection{Correctness}

The next part of Table \ref{tbl:evaluation} shows the percentage of points on non-outside patches which have been correctly oriented by our algorithm with respect to the ground truth orientations determined using given correspondences of points to known scanner positions.
We list the correctness after different phases of our algorithm.
Phase 1 is after the initial orientation by path tracing (Section \ref{subsec:pathtracing}), before (1A) and after (1B) making normals consistent within surfaces.
Phase 2 is after the fa\c{c}ade correction step (Section \ref{subsec:facade}), again before (2A) and after (2B) ensuring consistency within surfaces.

For each dataset, the images at the bottom of the table show horizontal cross sections of the point clouds for the upper and lower stories.
For each story, the upper image shows the unlabeled input point cloud.
The middle image shows the classification into interior surfaces (green), exterior (blue), outside (yellow), and points that are not on patches (gray).
The lower image shows the final correctness (after phase 2B) of the normal orientation with correctly oriented points (green), incorrectly oriented points (red), outside (yellow), and not on patches (gray).

Figure \ref{fig:bounces} shows the effect of allowing multiple bounces in our path tracing approach. The images show the orientation correctness after the path tracing phase and before applying surface consistency. Increasing the number of ray bounces helps to correctly identify the orientation of patches which are strongly occluded by surrounding rooms.
An overview of the classification of outside area is shown in Figure \ref{fig:classification} (a) which shows large areas scanned through windows or from balconies of the building. Outside area is colored yellow.
The detail view in Figure \ref{fig:classification} (b) shows a cross section of the same building with the more fine grained classification with the same color scheme as in Table \ref{tbl:evaluation}.
An example for fa\c{c}ade patches which are initially oriented incorrectly by the path tracing phase is shown in Figure \ref{fig:facade} (a). After applying the correction as described in Section \ref{subsec:facade}, the normals are oriented correctly (Figure \ref{fig:facade} b).
A failure case of the fa\c{c}ade correction step is shown in Figure \ref{fig:facade_fail}. The surface highlighted red was incorrectly oriented since the opened, almost parallel door next to it was interpreted as a wall surface. Note that this example is taken from Dataset 4 which explains the decreased final correctness as shown in Table \ref{tbl:evaluation}.

\subsection{Runtime}

Below the correctness percentages, Table \ref{tbl:evaluation} also lists the runtime of the individual phases of our algorithm as described above.
Clearly, the path tracing phase 1A takes more time than the single-bounce fa\c{c}ade correction phase 2A.
Also, the surface consistency correction 1B and 2B have similar runtimes since they are the same operation performed after phases 1A and 2A, respectively.
Even in case of the largest dataset (Dataset 4), the total runtime of the core normal orientation approach takes well below 10 seconds.
We are using the NVIDIA OptiX framework \cite{OptiX} for GPU-accelerated ray tracing against the set of patches which makes the actual ray tracing part a minor part of the total runtime requirements.
By far the largest contributor to the overall runtime of our approach is the plane detection for which we currently use a RANSAC implementation in the CGAL library.

%\subsection{Implementation details}

\section{Conclusion and future work}
\label{sec:conclusion}

We have presented a fast and fully automatic approach for orienting normals of the main building structures in multi-room, multi-story indoor point cloud measurements. The input to our algorithm are unstructured point clouds without any additional information such as scanner positions. Using a path tracing approach, we first classify points as interior, exterior, and outside surfaces, and estimate an initial orientation of all non-outside surfaces. In a second phase, we correct the orientation of fa\c{c}ade parts which may be incorrectly oriented in the first phase. Additionally, we perform a voting step for consistently orienting normals within surfaces after each phase.
We evaluated our approach on multiple, real-world datasets with respect to orientation correctness and runtime. The resulting, automatically estimated orientation information can greatly facilitate or enable tasks such as visualization or reconstruction of building models which rely on correctly oriented surface normals.

While the core of our algorithm provides fast processing of even larger datasets, the overall runtime is strongly dominated by the plane detection. One direction for future work is the evaluation of either different plane detection methods or alternatives for fast patch generation.
Also, the fa\c{c}ade correction step sometimes incorrectly interprets surfaces as boundaries of rooms and thus performs incorrect flipping of already correct normals. A more sophisticated interpretation of the surfaces may thus be a worthwhile direction for future research.

\section*{Acknowledgments}

{
This work was supported by the DFG projects KL 1142/11-1 (DFG Research Unit FOR 2535 Anticipating Human Behavior) and KL 1142/9-2 (DFG Research Unit FOR 1505 Mapping on Demand).
}

%%%%%%%%%%%% Supplementary Methods %%%%%%%%%%%%
%\footnotesize
%\section*{Methods}

%%%%%%%%%%%%% Acknowledgements %%%%%%%%%%%%%
%\footnotesize
%\section*{Acknowledgements}

%%%%%%%%%%%%%%   Bibliography   %%%%%%%%%%%%%%
\normalsize
\bibliography{references}

%%%%%%%%%%%%  Supplementary Figures  %%%%%%%%%%%%
%\clearpage

%%%%%%%%%%%%%%%%   End   %%%%%%%%%%%%%%%%
%\end{multicols}  % Method B for two-column formatting (doesn't play well with line numbers), comment out if using method A
\end{document}